# The first high precision differential abundance analysis of Extremely Metal Poor stars

Henrique Reggiani[1], Jorge Meléndez[1], David Yong[2], Ivan Ramírez[3], and Martin Asplund[2]

[1] Universidade de São Paulo, Insituto de Astronomia, Geofísica e Ciências Atmosféricas, IAG, Departamento de Astronomia, Rua do Matão 1226, Cidade Universitária, 05508-900, SP, Brazil. e-mail: hreggiani@gmail.com
[2] The Australian National University, Research School of Astronomy and Astrophysics, Weston, ACT 2611, Australia.
[3] University of Texas at Austin, McDonald Observatory and Department of Astronomy, TX, USA

**ABSTRACT**

*Context.* Studies of extremely metal-poor stars indicate that chemical abundance ratios [X/Fe] have an rms scatter as low as 0.05 dex (12 %). It remains unclear whether this reflects observational uncertainties or intrinsic astrophysical scatter arising from physical conditions in the ISM at early times.

*Aims.* Measure differential chemical abundance ratios in extremely metal-poor stars to investigate the limits of precision and to understand whether cosmic scatter or observational errors are dominant.

*Methods.* We used high resolution ($R \sim 95,000$) and high S/N (S/N = 700 at 5000Å) HIRES/Keck spectra, to determine high precision differential abundances between two extremely metal-poor stars through a line-by-line differential approach. We determined stellar parameters for the star G64-37 with respect to the standard star G64-12. We performed EW measurements for the two stars for the lines recognized in both stars and performed spectral synthesis to study the carbon abundances.

*Results.* The differential approach allowed us to obtain errors of $\sigma(T_{\rm eff})$ = 27 K, $\sigma(\log g)$ = 0.06 dex, $\sigma$([Fe/H])= 0.02 dex and $\sigma(v_t)$ = 0.06 kms$^{-1}$. We estimated relative chemical abundances with a precision as low as $\sigma$([X/Fe]) $\approx$ 0.01 dex. The small uncertainties demonstrate that there are genuine abundance differences larger than the measurement errors. The observed Li difference can not be explained by the difference in mass, because the less massive star has more Li.

*Conclusions.* It is possible to achieve an abundance precision around $\approx 0.01 - 0.05$ dex for extremely metal-poor stars, opening new windows on the study of the early chemical evolution of the Galaxy.

## 1. Introduction

Extremely metal-poor stars (EMP, i.e. stars with [Fe/H] $< -3$) are relics of the early universe and can provide us with precious clues about the chemical evolution and formation of the Galaxy. These objects arguably offer the most powerful insights into the evolution, nucleosynthetic yields, and properties of the first supernovae (Audouze & Silk 1995; Ryan et al. 1996; Shigeyama & Tsujimoto 1998; Chieffi & Limongi 2002; Umeda & Nomoto 2002).

The most accurate abundance measurements in EMP stars come from Cayrel et al. (2004) and Arnone et al. (2005) with errors for [X/Fe] as low as 0.05 dex. A key open issue is whether the observed scatter in abundance ratios reflects genuine cosmic scatter or measurement uncertainties. Higher precision abundance studies of EMP stars are needed to clarify this issue, but such measurements are challenging as they require long exposures using 8-m class telescopes to obtain high resolution and high S/N data. To improve our precision we employed the differential technique in our analysis. Recently, the differential technique in twin stars, meaning stars with similar stellar parameters, made it possible to considerably improve the precision achieved in spectroscopic studies because many error sources, such as imprecise $\log(gf)$ values, largely cancel out, allowing a much better precision in the determination of relative stellar parameters and abundances. Studies with this technique have been used to recognize planet signatures on the chemical composition of stars (Meléndez et al. 2009; Ramírez et al. 2009; Tucci Maia et al. 2014; Biazzo et al. 2015), stellar evolution effects (Monroe et al. 2013; Tucci Maia et al. 2015), chemical evolution in the solar neighborhood (Nissen 2015), abundance anomalies in globular clusters (Yong et al. 2013), and distinct populations in the metal-rich halo (Nissen & Schuster 2010).

Here we explore, for the first time, the chemical composition of two EMP turn-off stars through a strictly differential analysis, achieving an unprecedent precision (0.01 dex) for a few of the analyzed species.

## 2. Observations and Data Reduction

Spectra of G64-12 and G64-37 were obtained with HIRES, the High Resolution Echelle Spectrometer (Vogt et al. 1994), on the Keck 10 m telescope at Mauna Kea. The star G64-12 was observed on June/16/2005 and G64-37 on Jan/19/2006. The observations were performed with the same setup using the slit E4 (0.4"x7"), resulting in a resolving power of $R \sim 95,000$, with a S/N = 700 at 5000Å and S/N= 900 around the Li 6707Å line. The spectra have a wavelength coverage ranging from $\sim$ 3900Å to 8300Å.

The orders were extracted using MAKEE[1], specially written to reduce HIRES spectra. We performed the Doppler correction and continuum normalization via IRAF.

## 3. Analysis

We used a line-by-line differential approach to obtain stellar parameters and chemical abundances, as described in our previous works (e.g. Meléndez et al. 2012; Yong et al. 2013; Ramírez et al. 2015). The 2014 version of the LTE analysis code MOOG (Sneden 1973) was employed, with Castelli et. al. (1997) atmospheric models.

The linelist was created inspecting each feature to verify that each chosen line could be measured on both spectra. The $\log(gf)$ values and energy levels are from VALD (Vienna Atomic Line Database). The Fe I lines were updated using data from Den Hartog et al. (2014) and transition probabilities for the Fe II lines are from Meléndez & Barbuy (2009). The Ti II values were updated using Lawler et al. (2013). We note that the choice of $\log(gf)$ values is inconsequential in a differential analysis.

The equivalent widths (EW) were measured by hand with the *splot* task in IRAF, using Gaussian profile fits. In order to determine the local continuum we compared each line in the two stars by overplotting the spectra in a 6Å window.

The complete linelist, including the EW for both objects, is presented in Table 3.

G64-12 is used as the standard star for the analysis, with the following stellar parameters: $T_{\rm eff}$=6463 K from the IRFM (Meléndez et al. 2010), log $g$=4.26 dex from the absolute magnitude (Nissen et al 2007)[2] and, using our EW, we obtained [Fe/H]=−3.20 dex and $v_t$=1.65 kms$^{-1}$. Then, we employed a strictly line-by-line differential approach to obtain the stellar parameters of G64-37. Using as reference the Fe I and Fe II abundances from G64-12 we determined $T_{\rm eff}$=6570 K through differential excitation equilibrium (Fig. 1), consistent with the IRFM value ($T_{\rm eff}$=6583 ± 50 K, Meléndez et al. (2010)). We obtained a log $g$=4.40 dex through differential ionization equilibrium, consistent with Nissen et al (2007) (log $g$=4.24 ± 0.15). We obtained $v_t$=1.74 kms$^{-1}$ by allowing no trend in the differential Fe I line abundances with reduced EW (Fig. 1), and found [Fe/H]=−3.00 dex. The errors for the atmospheric parameters are: $\sigma(T_{\rm eff})$ = 27 K, $\sigma$(log $g$) = 0.06 dex, $\sigma$([Fe/H])= 0.02 dex and $\sigma(v_t = 0.06)$ $kms^{-1}$. They include the degeneracy of stellar parameters and were determined strictly through a differential approach.

Once the stellar parameters of G64-37 were determined through the iron lines, we determined the abundance of the other elements recognized in both spectra: Li, O, Na, Mg, Al, Si, Ca, Sc, Ti, Cr, Mn, Fe, Co Ni, Zn, Sr and Ba. For Li, Mn, Co and Ba hyperfine splitting was accounted for. For Li we used the linelist described in Meléndez & Ramírez (2004). For Mn and Co we employed the linelists from Kurucz[3] and for Ba we employed the linelist from McWilliam (1998). We present the final differential abundances in Table 1, along with the errors from propagating the stellar parameter errors and the observational error. The total errors were calculated by quadratically adding both observational and systematic errors. We also show, in the last column of Table 1, the ratio between differential abundances and total errors. This column reveals there are genuine abundance differences, greater than $2\sigma$ significance for $\Delta$[X/H], for all elements (except oxygen and silicon) between the two stars.

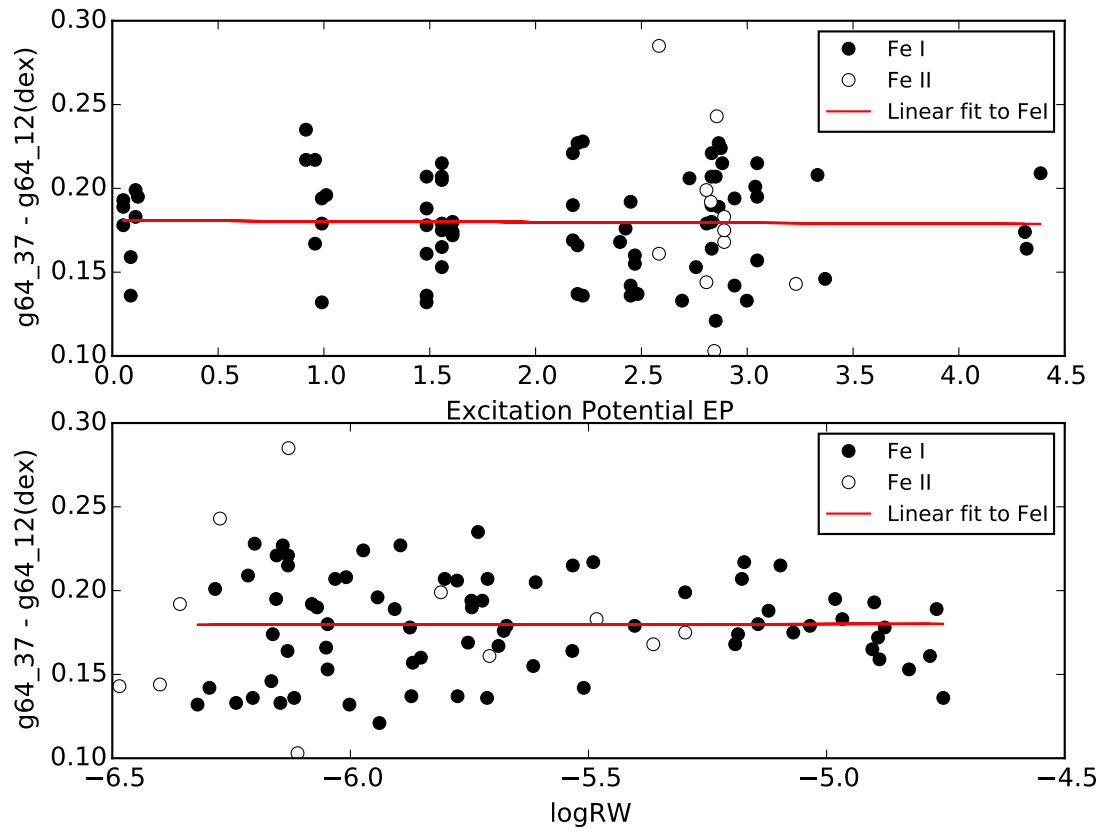

**Fig. 1.** Differential abundances versus lower excitation potential (top panel) and reduced equivalent widths (lower panel).

To demonstrate the importance of the differential technique in this work we have analyzed the [Mg/H] ratio for star G64-12 in a non differential way (classic analysis), achieving a much higher total error. The observational error ($\sigma/\sqrt{N}$) alone (0.059 dex) is higher than the total error obtained by using the differential analysis; when added to the parameters uncertainties (0.021 dex) the final error associated with the measurement is ≈ 0.083 dex, much higher than the 0.026 dex achieved using the differential technique.

We also present the differential abundance results relative to Fe ($\Delta$[X/Fe]). In this case the errors were derived considering how the errors for each stellar parameter behaves in relation to the same error in the iron differential abundance. After that we quadratically added the new parameters errors with the observational errors (defined as $\sigma/\sqrt{N}$, where N is the number of measured lines), presented in Table 1. We can see, through the significance of our results, shown in the forth column of Table 2, that working with [X/Fe] has decreased the confidence in the result of some elements, when compared to the results of [X/H] (Table 1). Eleven out of 17 species exhibit abundance differences (greater than 2-sigma significance) between the two stars for $\Delta$[X/Fe]. For the remaining six elements, the majority are heavy elements for which the total error is dominated by observational uncertainties arising from the small numbers of weak spectral lines, as can be seen in Table 3.

To further show the improvement that the differential technique offers, in Figure 2 we compare our errors with those obtained by Cayrel et al. (2004, Table 9) using a classical analysis. The dashed line represents the median value of the ratios between both errors, showing that our results are about four times more precise than the aforementioned work.

For carbon it was more appropriate to determine the abundances by spectral synthesis of the CH band. First, we estimated the Macro-turbulent ($V_{macro}$) velocity of the stars by visually fitting four different iron lines (3920.2Å, 4005.2Å, 4045.8Å, 4063.6Å, ). We determined $V_{macro}$ = $3.8 km.s^{-1}$ for G64-12 and $V_{macro}$ = $3.7 km.s^{-1}$ for G64-37.

We prepared a linelist, spanning from 4290Å to 4335Å, specifically for the carbon synthesis, with CH data from Masseron et al. (2014) along with atomic blends for the region from VALD. We synthesized, for each star, three different regions of the CH band, 4299Å to 4302Å, 4308Å to 4315Å and 4322Å to 4327Å. An example of a best fit for one of the regions, for star G64-12, can be seen in Figure 3. We averaged the abundance determination for the three regions and determined the abundance difference between the stars. We determined the parameters errors by synthesizing the three regions for each different parameter uncertainty.

[1] The package was created by T. A. Barlow and is freely available at http://www.astro.caltech.edu/ tb/makee/.

[2] The parallax is too uncertain, hence we adopted the photometric $M_V$ from Nissen et al (2007)

[3] http://kurucz.harvard.edu/linelists.html

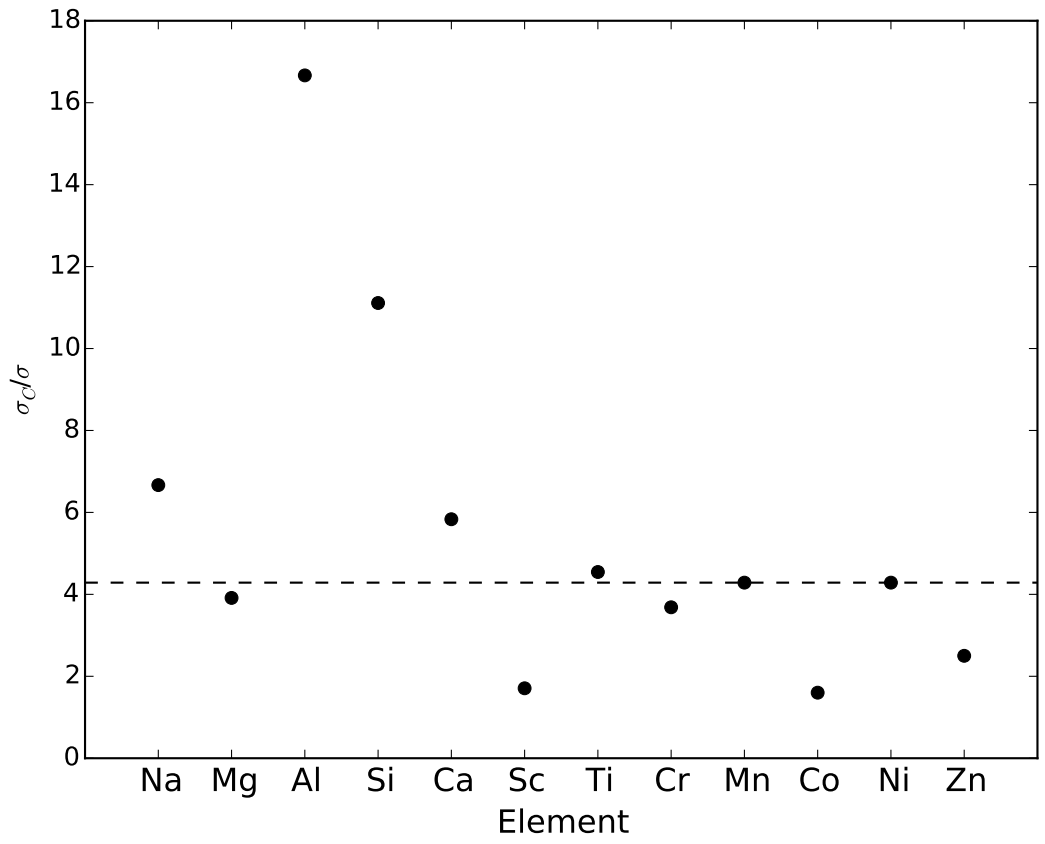

**Fig. 2.** The ratio between measurement errors from Cayrel et al. (2004)($\sigma_C$) and the errors obtained in this work ($\sigma$) for a number of elements.

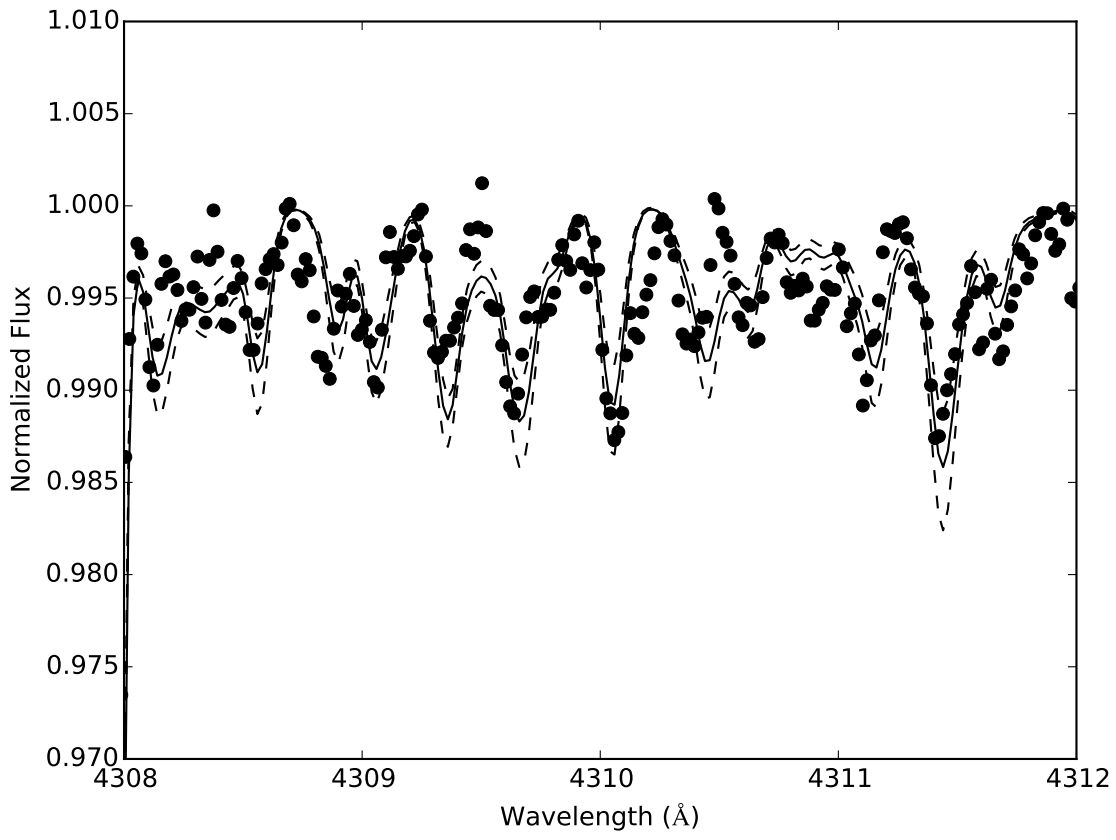

**Fig. 3.** Best fit of one of the regions synthesized for determining the carbon abundances (star G64-12). The dashed lines are a ± 0.1 dex difference in C abundance.

We also estimated ages and masses, using the q2 code (Ramírez et al. 2014). The code fits $Y^2$ isochrones (Yi et al. 2001; Kim et al. 2002) with the adopted stellar parameters. The method estimates the age and mass through a probability distribution approach, as described in Ramírez et al. (2013). For G64-12 we estimated an age of $14.0^{+0.6}_{-1.1}$ Gyr with a mass M=$0.76^{+0.01}_{-0.01}$ $M_\odot$. The best solution for star G64-37 is an age of $10.1^{+1.2}_{-2.1}$ Gyr with a mass M=$0.80^{+0.02}_{-0.02}$ $M_\odot$. The error bars represent the 68% confidence threshold.

It is important to stress that we derived the stellar ages through a probability density function (PDF) and obtained that star G64-12 is older than 12.9 Gyr with 68% certainty, and older than about 11.5 Gyr with 92% certainty. The probability of star G64-12 being as young as star G64-37 (10 Gyr) is as low as 0.3%. Star G64-37 is younger than 11.3 Gyrs with 68% certainty, and younger than 12.3 with 92% certainty.

**Table 2.** $\Delta$[X/Fe] differential abundances (G64-37 - G64-12).

| Species | $\Delta$[X/Fe] | Error | $\Delta$[X/Fe]/$\sigma$ | $\Delta$[X/Fe]$^C$ |
|---|---|---|---|---|
| Li I | −0.278 | 0.008 | 34.8 | ——— |
| C | 0.050 | 0.034 | 1.5 | 0.161 |
| O I | −0.173 | 0.065 | 2.7 | ——— |
| Na I | −0.125 | 0.015 | 8.3 | ——— |
| Mg I | −0.108 | 0.023 | 4.7 | −0.112 |
| Al I | −0.102 | 0.006 | 17.0 | ——— |
| Si I | −0.133 | 0.009 | 14.8 | −0.131 |
| Ca I | −0.094 | 0.012 | 7.8 | −0.106 |
| Sc II | −0.013 | 0.041 | 0.3 | 0.003 |
| Ti | −0.046 | 0.011 | 4.3 | −0.052 |
| Cr I | 0.058 | 0.019 | 3.1 | 0.025 |
| Mn I | 0.104 | 0.021 | 4.9 | 0.057 |
| Co I | −0.048 | 0.050 | 1.0 | 0.021 |
| Ni I | 0.013 | 0.021 | 0.7 | −0.007 |
| Zn I | −0.053 | 0.040 | 1.3 | 0.029 |
| Sr II | −0.024 | 0.022 | 1.1 | 0.048 |
| Ba II | −0.294 | 0.019 | 15.5 | ——— |

**Notes.** [(C)] Data corrected for Galactic chemical evolution.

The masses of both stars were also derived trough a PDF and we obtain that star G64-12 is less massive than 0.77 $M_\odot$ with 68% certainty and less massive than 0.78 $M_\odot$ with 92% certainty. Star G64-37 is more massive than 0.78 $M_\odot$ with 68% certainty and more massive than 0.77 $M_\odot$ with 92% certainty. The chance of star G64-12 being as massive as G64-37 is only about 4.5%.

Based on our PDF we can say that star G64-12 is older and less massive than star G64-37 with a very high degree of confidence.

Notice that the difference in age between our pair is similar to the difference in age between "low alpha" and "high alpha" halo stars at [Fe/H] $> -2$ (Schuster et al. 2012).

We checked both our stellar parameters and abundances results using q2 code, using MARCS model atmospheres (Gustafsson et al. 2008)) and the 2014 version of MOOG to compute the curves of growth, and obtained consistent results.

## 4. Discussion and Conclusions

In Figure 4 we show our differential abundances. This figure demonstrates that the differential technique is capable of revealing subtle differences in the abundance pattern of metal-poor stars, due to the small errors of $\approx$ 0.01 – 0.02 dex. The precision achieved shows that the pair G64-12/G64-37 have distinct abundance patterns. To compare our results we have searched the literature for works that analyzed both stars and have similar S/N and resolution as ours. We found a work from Nissen et al (2007) and they measured $\Delta$[Zn/H]= $0.19 \pm 0.20$, in good agreement with our results. Fabbian et al. (2009) also found similar stellar parameters, $\Delta$[C/H]=+0.04 ± 0.21 and $\Delta$[O/H]=−0.03 ± 0.21. The difference in carbon abundances might be due to the different techniques used for the determinations, while we synthesized CH molecular bands, Fabbian et al. (2009) measured EW for CI lines (not available in our spectral coverage), but the values are consistent within the analysis errors. The oxygen abundance agrees with our data, within the errors. Our study, using high-quality observation demonstrates that it is possible to

**Table 1.** Relative Abundances (G64-37 minus G64-12) and associated uncertainties due to errors in stellar parameters and observations.

| Species | Δ[X/H] | $\Delta T_{\rm eff}$ +26 K | Δlog $g$ +0.06 dex | $\Delta v_t$ +0.06 $kms^{-1}$ | Δ[Fe/H] +0.02 dex | $Param^a$ | $Obs^b$ | $Total^c$ | Δ[X/H]/σ |
|---|---|---|---|---|---|---|---|---|---|
| | ($dex$) | ($dex$) | ($dex$) | ($dex$) | ($dex$) | ($dex$) | ($dex$) | ($dex$) | |
| Li I | −0.098 | 0.020 | −0.001 | 0.000 | 0.000 | 0.020 | 0.006 | 0.021 | 4.7 |
| C | 0.230 | 0.020 | 0.010 | 0.020 | 0.010 | 0.030 | 0.02 | 0.037 | 6.2 |
| O I | 0.007 | −0.020 | 0.019 | 0.000 | 0.000 | 0.028 | 0.045 | 0.053 | 0.1 |
| Na I | 0.055 | 0.018 | −0.001 | −0.002 | −0.001 | 0.018 | 0.014 | 0.023 | 2.4 |
| Mg I | 0.072 | 0.014 | −0.003 | −0.004 | 0.000 | 0.015 | 0.021 | 0.026 | 2.8 |
| Al I | 0.078 | 0.022 | −0.001 | −0.002 | −0.001 | 0.022 | 0.005 | 0.023 | 3.4 |
| Si I | 0.047 | 0.022 | −0.001 | −0.008 | −0.001 | 0.023 | 0.007 | 0.024 | 1.9 |
| Ca I | 0.086 | 0.016 | −0.001 | −0.002 | 0.000 | 0.016 | 0.010 | 0.019 | 4.5 |
| Sc II | 0.167 | 0.013 | 0.020 | 0.000 | 0.000 | 0.024 | 0.034 | 0.042 | 4.0 |
| Ti I | 0.129 | 0.024 | −0.001 | 0.000 | −0.001 | 0.024 | 0.011 | 0.026 | 5.0 |
| Ti II | 0.155 | 0.010 | 0.019 | −0.001 | 0.000 | 0.021 | 0.007 | 0.023 | 6.7 |
| Cr I | 0.238 | 0.025 | −0.001 | −0.001 | −0.001 | 0.025 | 0.018 | 0.031 | 7.7 |
| Mn I | 0.284 | 0.028 | −0.001 | −0.001 | −0.001 | 0.028 | 0.020 | 0.034 | 8.4 |
| Fe I | 0.180 | 0.022 | −0.001 | −0.003 | 0.000 | 0.022 | 0.003 | 0.022 | 8.3 |
| Fe II | 0.181 | 0.004 | 0.020 | −0.001 | 0.000 | 0.020 | 0.015 | 0.025 | 7.2 |
| Co I | 0.132 | 0.026 | 0.000 | 0.000 | −0.001 | 0.026 | 0.050 | 0.056 | 2.4 |
| Ni I | 0.193 | 0.013 | 0.001 | 0.000 | 0.000 | 0.013 | 0.018 | 0.022 | 8.8 |
| Zn I | 0.127 | 0.014 | 0.004 | −0.001 | −0.001 | 0.015 | 0.039 | 0.042 | 3.0 |
| Sr II | 0.156 | 0.016 | 0.018 | −0.010 | 0.000 | 0.026 | 0.005 | 0.027 | 5.8 |
| Ba II | −0.114 | 0.018 | 0.017 | −0.001 | 0.000 | 0.025 | 0.004 | 0.025 | 4.6 |

**Notes.** [a] Errors due to stellar parameters. [b] Observational error, s.e=$\sigma/\sqrt{N}$. [c] Total error, quantified as the quadratic sum of the stellar parameters errors and the observational error.

study, for example, the separation of the halo population via the abundance pattern of alpha elements Mg, Si and Ti, shown to exist by Nissen & Schuster (2010) in more metal-rich halo stars.

The differential abundances presented in Table 1 are indicative that these two stars belong to two different populations, as there is a significant difference on the abundances of all analyzed elements. In the last column of Table 1 we show the significance of our results and it can be seen that all of our results can be trusted with over 2$\sigma$ confidence, with the exception of oxygen.

Analyzing the $\alpha$-elements one can point how the errors must be small to separate the stars via differential abundances: Δ[Ti/H]=0.142 ± 0.035, Δ[O/H]=0.007 ± 0.053, Δ[Mg/H]=0.072 ± 0.026 and Δ[Si/H]=0.047 ± 0.024 are very small. Thus, to distinguish a clear difference we have to achieve errors on the order of (0.01 − 0.02 dex). There is a small abundance difference between the stars which indicate that they might belong to distinct halo populations.

As in Nissen & Schuster (2010) and Ramírez et al. (2012), we can analyze the possibility of distinct halo populations through [$\alpha$/Fe] ratios. As can be seen in Table 2, when compared to iron, the differential abundances between these stars are more prominent (Δ[O/Fe]= −0.173 ± 0.065, Δ[Mg/Fe]= −0.108 ± 0.023, Δ[Si/Fe]= −0.133±0.009 and Δ[Ti/Fe]= −0.046±0.011), also indicating that they belong to distinct halo populations. It is important to stress that for all the $\alpha$-elements the significance of our results are all above 2$\sigma$, including for [O/Fe]. With this data we find that G64-37, the younger halo star, has lower [$\alpha$/Fe], which is in agreement with the results of Schuster et al. (2012).

In order to exclude differences that might arise from Galactic chemical evolution (GCE) we performed linear regressions to the data published in Bonifacio et al. (2009), who performed abundance analysis for stars with similar stellar parameters as here, but in a wider range of metallicities. Then, we corrected our [X/Fe] ratios for the predicted ratio of the linear regression. We present the corrected differential abundances, Δ[X/Fe]$^C$, for trends in galactic chemical evolution, in the last column of Table 2. We notice that the galactic chemical evolution corrections are within the error bars of our results and do not change our interpretation, as can also be seen in Figure 4.

We also estimated the velocity components for the two stars, using an estimated distance from the absolute magnitude by Nissen et al (2007), proper motion data from van Leeuwen (2007) and radial velocity from Latham et al. (2002). For star G64-12 we obtained $U_{LSR}$=21 km/s, $V_{LSR}$=−352 km/s and $W_{LSR}$=−400 km/s and for star G64-37 $U_{LSR}$=231 km/s, $V_{LSR}$=−369 km/s and $W_{LSR}$=−77 km/s.[4] We found that both stars have extreme kinematics, falling outside Figure 3 from Nissen & Schuster (2010). However, it is important to point that their study present stars with metallicities [Fe/H]> −1.5, much higher than the stars in this work.

With the small errors achieved, it is also possible to revisit the Li plateau (Spite & Spite 1982). Meléndez et al. (2010) demonstrated the existence of two plateaus, with a break at [Fe/H]≈ −2.5. With an error of ∼ 0.021 on our Li differential abundance it will be possible to study a larger sample of stars and determine with higher precision where is the break of the Li plateau. As the two stars have metallicities that place them on the same plateau we can compare the absolute differential abundance with the scatter found by Meléndez et al. (2010). The differential Δ[Li/H] abundance found in this study (0.098 dex) is higher than the average scatter (0.04 dex) previously found among stars in that range ([Fe/H]< −2.5).

In Meléndez et al. (2010) the difference in Li was argued to be due to the differences in mass between the stars, as stars with lower masses deplete more lithium (Richard et al. 2005). However, the pair studied here behave unexpectedly as the more metal-poor, oldest and less massive star seems to have a higher Li content. In order to check the result we also performed NLTE abundance corrections (Lind et al. 2009) and arrived at a differential NLTE abundance of Δ[Li/H] = −0.10 dex, showing the consistency of our results. Presently there are only Li diffusion models for [Fe/H]=−2 (Richard et al. 2005). It would be important to extend these models to lower metallicities to test against our high precision Li abundances.

[4] The Hipparcos parallaxes are too uncertain. Better velocity components will be obtained once GAIA results are released.

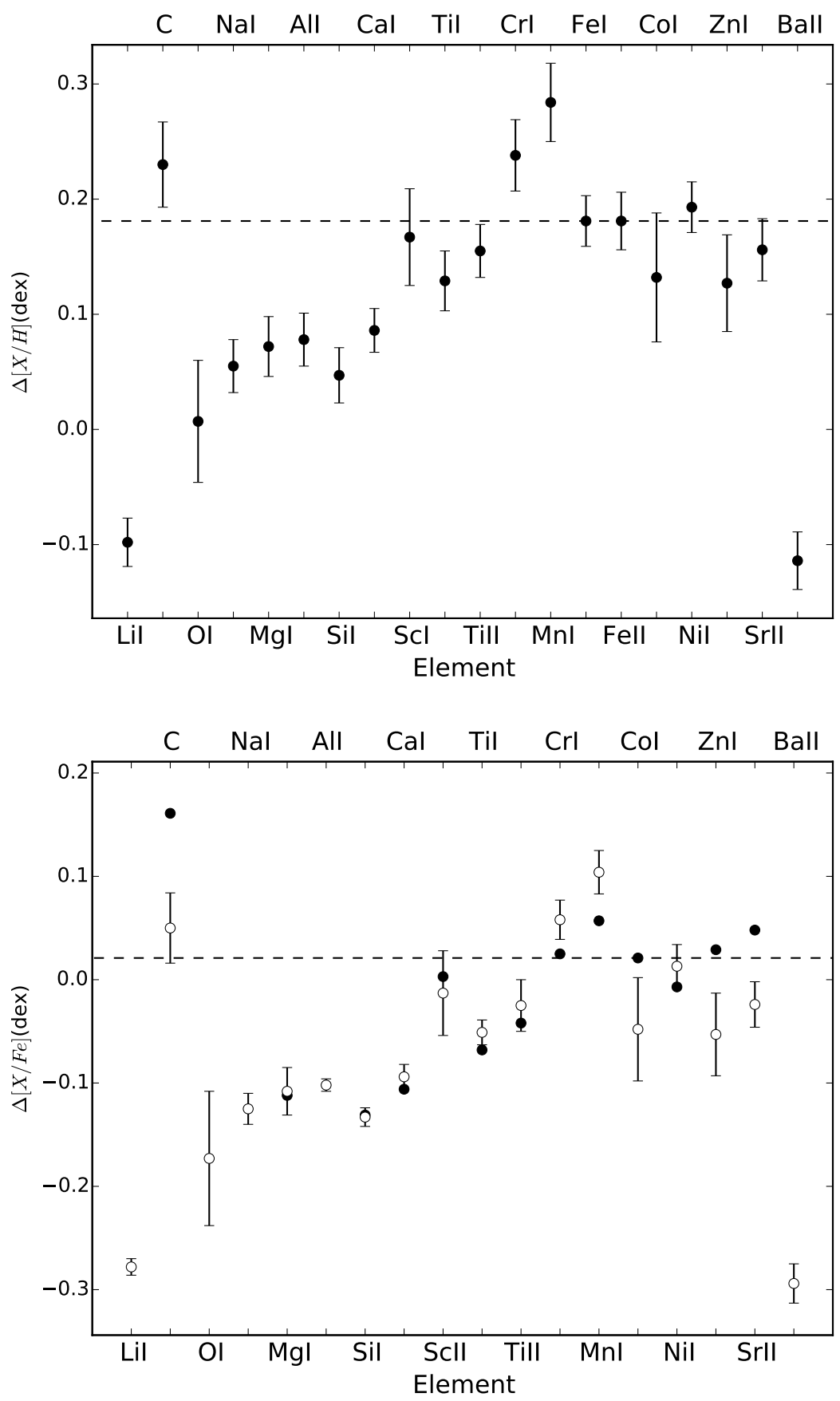

**Fig. 4.** Top panel: Δ[X/H] abundances. Lower panel: Δ[X/Fe] abundances. Filled circles are the values corrected for Galactic chemical evolution, while the open circles represent the uncorrected abundances (G64-37 - G64-12).

The results here presented illustrate how a differential study can help indicate if lithium is, in fact, being depleted in stars or if physics beyond the primordial nucleosynthesis model is necessary (Fields et al. 2014).

Even after GCE corrections, there remain clear abundance differences even among chemical elements produced via similar processes. For example, oxygen is more enhanced than carbon in G64-12, as is also the case of barium and strontium. The difference in the abundance patterns of these stars can give us important information on the environment in which these two stars formed and on the supernovae that enriched them.

We attempted to determine possible supernovae progenitors for our stars. In order to do so we have employed the STARFIT[5] code (Chan et al. 2015) with the absolute abundances calculated for our standard star (G64-12) and the absolute values for the standard star plus the differential abundances (Table 1), to study a possible progenitor for star G64-37.

We found no extreme difference between the possible polluting supernovae. The results from STARFIT indicate that the star G64-12 had a progenitor with mass M= $18 M_\odot$, log(*mixing*)= −1.0 dex and a remnant of $3.9 M_\odot$. The results of G64-37 implied a supernovae with M= $11 M_\odot$, log(*mixing*)= −1.6 dex and a remnant of $1.6 M_\odot$.

Our study demonstrates that the advent of precision spectroscopy can open new windows on the study of the early Galaxy, supernovae yields and chemical evolution of the Galaxy. With a larger sample of very metal-poor stars we will be able to assess additional issues such as cosmic scatter in the Galactic halo, and how the first supernovae enriched our Galaxy.

*Acknowledgements.* HR thanks a CAPES fellowship. JM thanks support by FAPESP (2010/50930-6 and 2012/24392-2). DY was supported through an Australian Research Council Future Fellowship (FT140100554). MA has been supported by the Australian Research Council grant (FL110100012).

[5] http://starfit.org/

**Table 3.** Linelist used for the abundances determinations.

| Wavelength (Å) | Species | EP (eV) | log($gf$) (dex) | G64-12(EW) (mÅ) | G64-37(EW) (mÅ) |
|---|---|---|---|---|---|
| 3886.282 | 26.0 | 0.052 | −1.076 | 61.40 | 66.20 |
| 3887.048 | 26.0 | 0.915 | −1.144 | 21.20 | 26.10 |
| 3895.656 | 26.0 | 0.110 | −1.670 | 37.70 | 42.00 |
| 3899.707 | 26.0 | 0.087 | −1.531 | 47.10 | 50.40 |
| 3902.946 | 26.0 | 1.557 | −0.466 | 31.30 | 36.00 |
| 3906.480 | 26.0 | 0.110 | −2.243 | 16.60 | 19.70 |
| 3917.181 | 26.0 | 0.990 | −2.155 | 3.60 | 3.90 |
| 3920.258 | 26.0 | 0.121 | −1.746 | 36.10 | 40.90 |
| 3922.912 | 26.0 | 0.052 | −1.651 | 44.50 | 49.40 |
| 3997.392 | 26.0 | 2.727 | −0.479 | 5.00 | 6.70 |
| 3998.053 | 26.0 | 2.692 | −0.910 | 2.00 | 2.30 |
| 4005.242 | 26.0 | 1.557 | −0.610 | 26.00 | 32.00 |
| 4009.713 | 26.0 | 2.223 | −1.252 | 2.20 | 2.50 |
| 4014.531 | 26.0 | 3.047 | −0.587 | 2.10 | 2.80 |
| 4021.867 | 26.0 | 2.758 | −0.729 | 3.00 | 3.60 |
| 4045.812 | 26.0 | 1.485 | 0.280 | 67.50 | 71.20 |
| 4063.594 | 26.0 | 1.557 | 0.062 | 56.20 | 60.50 |
| 4132.058 | 26.0 | 1.608 | −0.675 | 23.10 | 26.90 |
| 4134.678 | 26.0 | 2.831 | −0.649 | 2.90 | 3.70 |
| 4143.415 | 26.0 | 3.047 | −0.204 | 4.60 | 5.60 |
| 4143.868 | 26.0 | 1.557 | −0.511 | 30.80 | 35.30 |
| 4181.755 | 26.0 | 2.831 | −0.371 | 4.90 | 6.60 |
| 4187.039 | 26.0 | 2.449 | −0.548 | 7.10 | 8.10 |
| 4187.795 | 26.0 | 2.425 | −0.554 | 7.10 | 8.80 |
| 4191.431 | 26.0 | 2.469 | −0.666 | 4.90 | 5.90 |
| 4199.095 | 26.0 | 3.047 | 0.155 | 9.10 | 12.30 |
| 4202.029 | 26.0 | 1.485 | −0.708 | 26.90 | 31.70 |
| 4222.213 | 26.0 | 2.449 | −0.967 | 2.70 | 3.50 |
| 4233.603 | 26.0 | 2.482 | −0.604 | 6.20 | 7.10 |
| 4247.426 | 26.0 | 3.368 | −0.239 | 2.40 | 2.90 |
| 4250.119 | 26.0 | 2.469 | −0.405 | 8.70 | 10.30 |
| 4250.787 | 26.0 | 1.557 | −0.714 | 23.00 | 28.20 |
| 4260.474 | 26.0 | 2.399 | 0.109 | 23.30 | 27.40 |
| 4271.154 | 26.0 | 2.449 | −0.349 | 11.50 | 13.20 |
| 4271.761 | 26.0 | 1.485 | −0.164 | 51.20 | 56.60 |
| 4282.403 | 26.0 | 2.176 | −0.779 | 6.10 | 7.70 |
| 4315.085 | 26.0 | 2.198 | −0.965 | 5.10 | 5.80 |
| 4325.762 | 26.0 | 1.608 | 0.006 | 50.30 | 55.50 |
| 4383.545 | 26.0 | 1.485 | 0.200 | 67.40 | 72.20 |
| 4404.750 | 26.0 | 1.557 | −0.142 | 50.10 | 54.90 |
| 4415.123 | 26.0 | 1.608 | −0.615 | 27.20 | 31.70 |
| 4427.310 | 26.0 | 0.052 | −2.924 | 5.10 | 5.90 |
| 4442.339 | 26.0 | 2.198 | −1.255 | 2.30 | 3.20 |
| 4447.717 | 26.0 | 2.223 | −1.342 | 2.00 | 2.80 |
| 4459.118 | 26.0 | 2.176 | −1.279 | 2.40 | 3.30 |
| 4461.653 | 26.0 | 0.087 | −3.210 | 3.20 | 3.40 |
| 4466.552 | 26.0 | 2.831 | −0.600 | 2.90 | 3.80 |
| 4494.563 | 26.0 | 2.198 | −1.136 | 3.30 | 4.00 |
| 4528.614 | 26.0 | 2.176 | −0.822 | 6.60 | 8.00 |
| 4602.941 | 26.0 | 1.485 | −2.209 | 2.00 | 2.20 |
| 4871.318 | 26.0 | 2.865 | −0.363 | 4.40 | 6.20 |
| 4872.138 | 26.0 | 2.882 | −0.567 | 2.60 | 3.60 |
| 4890.755 | 26.0 | 2.875 | −0.394 | 3.70 | 5.20 |
| 4891.492 | 26.0 | 2.851 | −0.112 | 7.10 | 9.50 |
| 4918.994 | 26.0 | 2.865 | −0.342 | 4.70 | 6.10 |
| 4920.503 | 26.0 | 2.832 | 0.068 | 11.90 | 14.40 |
| 4957.299 | 26.0 | 2.851 | −0.408 | 5.10 | 5.70 |
| 4957.597 | 26.0 | 2.808 | 0.233 | 15.90 | 19.60 |
| 5006.119 | 26.0 | 2.832 | −0.638 | 2.50 | 3.50 |
| 5139.463 | 26.0 | 2.940 | −0.509 | 2.20 | 2.60 |

**Table 3.** Continued.

| Wavelength (Å) | Species | EP (eV) | log($gf$) (dex) | G64-12(EW) (mÅ) | G64-37(EW) (mÅ) |
|---|---|---|---|---|---|
| 5171.596 | 26.0 | 1.485 | −1.793 | 3.70 | 4.80 |
| 5191.455 | 26.0 | 3.038 | −0.551 | 2.00 | 2.70 |
| 5192.344 | 26.0 | 2.998 | −0.421 | 3.20 | 3.70 |
| 5227.190 | 26.0 | 1.557 | −1.228 | 10.10 | 12.80 |
| 5232.940 | 26.0 | 2.940 | −0.058 | 7.20 | 9.40 |
| 5371.490 | 26.0 | 0.958 | −1.645 | 13.80 | 17.40 |
| 5383.369 | 26.0 | 4.312 | 0.645 | 2.80 | 3.70 |
| 5397.128 | 26.0 | 0.915 | −1.993 | 7.50 | 10.00 |
| 5405.775 | 26.0 | 0.990 | −1.844 | 9.70 | 11.50 |
| 5415.199 | 26.0 | 4.386 | 0.642 | 2.30 | 3.30 |
| 5424.068 | 26.0 | 4.320 | 0.520 | 3.10 | 4.00 |
| 5429.697 | 26.0 | 0.958 | −1.879 | 9.60 | 11.10 |
| 5434.524 | 26.0 | 1.011 | −2.122 | 5.00 | 6.20 |
| 5446.917 | 26.0 | 0.990 | −1.914 | 8.40 | 10.30 |
| 5615.644 | 26.0 | 3.332 | 0.050 | 4.00 | 5.50 |
| 4178.862 | 26.1 | 2.583 | −2.510 | 1.90 | 3.10 |
| 4233.172 | 26.1 | 2.583 | −1.970 | 6.80 | 8.30 |
| 4508.288 | 26.1 | 2.856 | −2.440 | 1.60 | 2.40 |
| 4520.224 | 26.1 | 2.807 | −2.650 | 1.50 | 1.80 |
| 4522.634 | 26.1 | 2.844 | −2.250 | 3.20 | 3.50 |
| 4555.893 | 26.1 | 2.828 | −2.400 | 1.50 | 2.00 |
| 4583.837 | 26.1 | 2.807 | −1.930 | 5.30 | 7.10 |
| 4923.927 | 26.1 | 2.891 | −1.260 | 12.80 | 16.20 |
| 5018.440 | 26.1 | 2.891 | −1.100 | 17.90 | 21.70 |
| 5169.033 | 26.1 | 2.891 | −1.000 | 21.40 | 26.10 |
| 5197.577 | 26.1 | 3.230 | −2.220 | 1.40 | 1.70 |
| 6707.820 | 3.0 | 0.000 | 0.167 | 24.10 | 16.60 |
| 7771.941 | 8.0 | 9.146 | 0.369 | 4.50 | 5.30 |
| 7774.161 | 8.0 | 9.146 | 0.223 | 3.80 | 3.70 |
| 5889.951 | 11.0 | 0.000 | 0.117 | 31.70 | 29.50 |
| 5895.924 | 11.0 | 0.000 | −0.184 | 19.00 | 19.70 |
| 4057.505 | 12.0 | 4.346 | −1.201 | 2.70 | 3.70 |
| 4167.271 | 12.0 | 4.346 | −1.004 | 4.70 | 4.40 |
| 4351.906 | 12.0 | 4.346 | −0.833 | 5.80 | 5.70 |
| 5167.321 | 12.0 | 2.709 | −1.030 | 45.40 | 47.50 |
| 5172.684 | 12.0 | 2.712 | −0.402 | 75.70 | 76.10 |
| 5183.604 | 12.0 | 2.717 | −0.180 | 89.30 | 89.80 |
| 5528.405 | 12.0 | 4.346 | −0.620 | 8.10 | 9.10 |
| 3944.006 | 13.0 | 0.000 | −0.623 | 17.80 | 18.00 |
| 3961.520 | 13.0 | 0.014 | −0.323 | 24.00 | 23.30 |
| 3905.523 | 14.0 | 1.909 | −0.743 | 55.80 | 54.70 |
| 4226.728 | 20.0 | 0.000 | 0.244 | 86.50 | 86.60 |
| 4283.011 | 20.0 | 1.886 | −0.292 | 5.80 | 6.00 |
| 4289.367 | 20.0 | 1.879 | −0.388 | 4.10 | 4.80 |
| 4318.652 | 20.0 | 1.899 | −0.295 | 5.10 | 6.00 |
| 4425.437 | 20.0 | 1.879 | −0.358 | 4.30 | 4.70 |
| 4435.679 | 20.0 | 1.886 | −0.517 | 3.30 | 3.30 |
| 4454.779 | 20.0 | 1.899 | 0.258 | 14.20 | 14.50 |
| 4455.887 | 20.0 | 1.899 | −0.414 | 3.00 | 3.60 |
| 5588.749 | 20.0 | 2.526 | 0.358 | 5.60 | 6.50 |
| 5594.462 | 20.0 | 2.523 | 0.097 | 3.70 | 3.70 |
| 5857.451 | 20.0 | 2.933 | 0.240 | 2.30 | 2.30 |
| 6122.217 | 20.0 | 1.886 | −0.386 | 4.60 | 5.40 |
| 6162.173 | 20.0 | 1.899 | −0.167 | 8.00 | 8.10 |
| 6439.075 | 20.0 | 2.526 | 0.394 | 6.90 | 7.10 |
| 4246.822 | 21.1 | 0.315 | 0.242 | 13.30 | 14.80 |
| 4314.083 | 21.1 | 0.618 | −0.096 | 3.30 | 4.50 |
| 4320.732 | 21.1 | 0.605 | −0.252 | 2.40 | 2.20 |
| 4374.457 | 21.1 | 0.618 | −0.418 | 1.70 | 1.90 |
| 4400.389 | 21.1 | 0.605 | −0.536 | 1.30 | 1.80 |

**Table 3.** Continued.

| Wavelength (Å) | Species | EP (eV) | log($gf$) (dex) | G64-12(EW) (mÅ) | G64-37(EW) (mÅ) |
|---|---|---|---|---|---|
| 3958.206 | 22.0 | 0.048 | −0.177 | 3.50 | 4.10 |
| 3989.759 | 22.0 | 0.021 | −0.198 | 3.40 | 3.80 |
| 3998.636 | 22.0 | 0.048 | −0.056 | 4.40 | 4.20 |
| 4305.908 | 22.0 | 0.848 | 0.510 | 3.30 | 3.60 |
| 4533.241 | 22.0 | 0.848 | 0.476 | 2.70 | 3.30 |
| 4534.776 | 22.0 | 0.836 | 0.280 | 2.00 | 2.20 |
| 4535.568 | 22.0 | 0.826 | 0.162 | 0.90 | 1.10 |
| 4981.731 | 22.0 | 0.848 | 0.504 | 3.30 | 3.70 |
| 4991.065 | 22.0 | 0.836 | 0.380 | 2.80 | 2.90 |
| 4999.503 | 22.0 | 0.826 | 0.250 | 2.10 | 2.10 |
| 3900.539 | 22.1 | 1.131 | −0.290 | 26.70 | 29.10 |
| 3913.461 | 22.1 | 1.116 | −0.360 | 23.70 | 27.00 |
| 4012.383 | 22.1 | 0.574 | −1.840 | 4.80 | 5.70 |
| 4028.338 | 22.1 | 1.892 | −0.920 | 1.80 | 2.30 |
| 4290.215 | 22.1 | 1.165 | −0.870 | 8.00 | 9.80 |
| 4300.042 | 22.1 | 1.180 | −0.460 | 17.70 | 19.20 |
| 4301.922 | 22.1 | 1.161 | −1.210 | 4.80 | 6.10 |
| 4312.860 | 22.1 | 1.180 | −1.120 | 5.70 | 6.40 |
| 4395.031 | 22.1 | 1.084 | −0.540 | 20.80 | 24.20 |
| 4399.765 | 22.1 | 1.237 | −1.190 | 3.80 | 4.10 |
| 4417.714 | 22.1 | 1.165 | −1.190 | 4.60 | 4.90 |
| 4443.801 | 22.1 | 1.080 | −0.710 | 15.70 | 17.30 |
| 4450.482 | 22.1 | 1.084 | −1.520 | 2.60 | 3.40 |
| 4468.507 | 22.1 | 1.131 | −0.600 | 16.80 | 19.00 |
| 4501.270 | 22.1 | 1.116 | −0.770 | 13.00 | 14.90 |
| 4533.960 | 22.1 | 1.237 | −0.530 | 14.60 | 16.60 |
| 4549.622 | 22.1 | 1.584 | −0.110 | 18.50 | 21.50 |
| 4563.757 | 22.1 | 1.221 | −0.690 | 10.30 | 11.80 |
| 4571.971 | 22.1 | 1.572 | −0.320 | 13.10 | 15.70 |
| 4589.947 | 22.1 | 1.237 | −2.940 | 1.80 | 1.80 |
| 4254.336 | 24.0 | 0.000 | −0.114 | 15.40 | 19.70 |
| 4274.797 | 24.0 | 0.000 | −0.231 | 12.60 | 16.90 |
| 4289.717 | 24.0 | 0.000 | −0.361 | 9.20 | 12.30 |
| 5206.037 | 24.0 | 0.941 | 0.019 | 4.30 | 6.80 |
| 5208.425 | 24.0 | 0.941 | 0.158 | 7.00 | 8.60 |
| 4030.753 | 25.0 | 0.000 | −0.470 | 5.00 | 6.80 |
| 4030.730 | 25.0 | 0.000 | −1.037 | 5.00 | 6.80 |
| 3995.269 | 27.0 | 0.923 | −2.026 | 3.40 | 4.10 |
| 4121.294 | 27.0 | 0.923 | −0.993 | 3.00 | 2.90 |
| 4025.101 | 28.0 | 4.088 | −1.343 | 1.50 | 2.20 |
| 4810.528 | 30.0 | 4.078 | −0.137 | 1.10 | 1.20 |
| 4077.709 | 38.1 | 0.000 | 0.167 | 44.40 | 47.50 |
| 4215.519 | 38.1 | 0.000 | −0.145 | 32.20 | 35.30 |
| 4554.000 | 56.1 | 0.000 | −1.447 | 6.60 | 4.10 |
| 4934.100 | 56.1 | 0.000 | −1.767 | 3.90 | 2.35 |